# Performance of Step Network Using Simulation Tool

Taskeen Zaidi and Nitya Nand Dwivedi

*Abstract*—Nowadays distributed computing approach has become very popular due to several advantages over the centralized computing approach as it also offers high performance computing at a very low cost. Each router implements some queuing mechanism for resources allocation in a best possible optimize manner and governs with packet transmission and buffer mechanism. In this paper, different types of queuing disciplines have been implemented for packet transmission when the bandwidth is allocated as well as packet dropping occurs due to buffer overflow. This gives result in latency in packet transmission, as the packet has to wait in a queue which is to be transmitted again. Some common queuing mechanisms are first in first out, priority queue and weighted fair queuing, etc. This targets simulation in heterogeneous environment through simulator tool to improve the quality of services by evaluating the performance of said queuing disciplines. This is demonstrated by interconnecting heterogeneous devices through step topology. In this paper, authors compared data packet, voice and video traffic by analyzing the performance based on packet dropped rate, delay variation, end to end delay and queuing delay and how the different queuing discipline effects the applications and utilization of network resources at the routers. Before evaluating the performance of the connected devices, a Unified Modeling Language class diagram is designed to represent the static model for evaluating the performance of step topology. Results are described by taking the various case studies.

## I. INTRODUCTION

IN the present scenario, distributed computing is widely adopted by many researchers for computation of parallel execution of tasks in optimum time period as it represents the autonomous collection of heterogeneous systems interconnected through heterogeneous network. The devices may be computer system, mobile system, laptop, tablet and other types of handheld devices. Various kinds of topologies are studied by distributed computing approach and according to Coulouris et al. [1], distributed system is an autonomous collection of heterogeneous devices communicating with the help of message passing technique. The characterizations of distributed system, system model, inter-process communication, web services, security issues in distributed system and designing of distributed systems are described by the authors. Hwang [2] has described architecture of various computer models, program behavior, architectural choices, scalability, programmability, performance issues related to parallel processing, designing high-performance computers, supporting software and applications for distributed and parallel computing. Network topologies are one major building block for data communication and Wahlisch [3] described how network entities are directly interconnected with each other, also explained how information is flowing from one device to another device. A structure build of node relations can be built on different layers resulting in a physical or logical topology, constructed while connecting devices by a physical medium. Data exchange on top of this structure can be arranged via the network and application layer creates a logical or overlay topology. The detail description about queue, scheduling technique, QoS in network, queue size, queuing delay and TCP window policy are well explained in [4]. Latha and Srivatsa [5] illustrated the goal of topological design of a computer communication network which is used to accomplish a specified performance at a minimal cost.

Kamalesh and Srivatsa [6] observed the assignment of node number in a computer communication network using heuristic approach. Assessment of the performance of different topological structure based on Ant Colony Optimization Algorithm using simulator is available in [7]. Modeling of parallel and distributed applications was a captivation of numerous research groups in the past due to increasing the importance of applications on mixed shared memory parallelism with message passing method [8]. The various aspects of UML and different versions of UML are released by OMG group [9-10].

The Unified Modeling Language user guide, the prime developers of the UML--Grady Booch, James Rumbaugh, and Ivar Jacobson have presented a tutorial to describe the essence of the language in a two-color format which is designed to facilitate the learning. Outset with a conceptual model of the UML, the book on UML is used to solve series of complex modeling problems across a variety of application domains by various researchers [11-12].

Object-oriented distributed architecture system through UML has been explained by Arora et al. [13].Various authors are using distribu ted computer system which has become very popular approach of computing as it delivers high-end performance at a low cost. In a distributed computing environment, autonomous computers are linked by means of a communication network, arranged in a geometrical way called network topology. A detailed study of network topologies is executed for the distributed computer systems. A most popular object-oriented modeling language adopted by OMG i.e. Unified Modeling Language (UML) is used for modeling the different network topologies. A comparative study for 2D Mesh, Torus, and Hypercube network topologies and their performance is also evaluated after designing the UML model for class, sequence, and activity diagrams [14]. It has the generation of networks, network layer architecture; kinds of topologies used in the networking, different types of communication styles, types of secure transmission and finally covers the aspects of wireless types of secure transmission. Data communication, networks, internet, protocols, signals, congestion control and techniques to improve QoS are well explained by the Frouzan [15]. A method using evolutionary structural optimization method has been designed [16] in which the quality of solution is improved by ignoring chain like sets of elements which are responsible for potential kinematic instabilities and local error estimators. It also refines the mesh, to obtain an accurate and stable solution. Zaidi and Saxena[17] have used the concept of graph theory to design a step topology for static interconnection of devices across distributed network. An object-oriented approach UML is used to design model for execution of task in critical section and represented via class and sequence view. Space complexity is also calculated and result is shown in the form of tables and graphs.

Distributed computing approach has become an essential part of many of the software companies. In [18], a well-known object-oriented Unified Modeling Language (UML) is used to create a model for the execution of the tasks across a distributed network for the faster execution of tasks on newly designed step topology. Static and dynamic observations of the execution of the tasks are characterized through UML class and state diagrams, respectively. To validate the proposed model the state diagram is converted into a Finite State Machine (FSM) and different test cases are generated across distributed network environment.

Ricart and Agrawala inferred that if all sites must donate permission by sending replies, then the release messages are not required, since a reply involves an implicit release. It reduced message complexity [19]. Using a series of benchmark, author [20] studied comparison of algorithms for structural topology optimization based on (i) the artificial material model and (ii) a special microstructure with square voids done in [20]. Zaidi and Saxena [21] also used OPNET to simulate the results for video conferencing and voice packets connected across step networks and it is observed that delay is negligible; a UML class model was also designed to represent the optimized network.

[1]**Taskeen Zaidi**, Assistant Professor, Department of Computer Science, SRM University, Lucknow (U.P.) 225003, India,e-mail:taskeenzaid867@gmail.com

[2]**Nitya Nand Dwivedi**, Research Scholar, Department of Computer Science, SRM University, Lucknow (U.P.) 225003 India, e-mail: nityananddwivedi29@gmail.com

### III. BACKGROUND
*A. Step Topology*
It is a new kind of topology called as step topology for static interconnection of handheld devices in step manner which may be desktop computer systems, laptops, mobile, etc. This topology is modification of bus topology by varying the steps which interconnect N numbers of computer systems. This topology works well if the link between two computers connected through bus topology fails. In this topology if individual node is busy then tasks can be executed on next node by using message passing technique and the devices can be connected in static as well as dynamically through adhoc network. A view of step topology through OPNET simulation tool is sketched in figure 1. The detailed narration about step topology can be found in [16].

*B. OPNET Modeler*

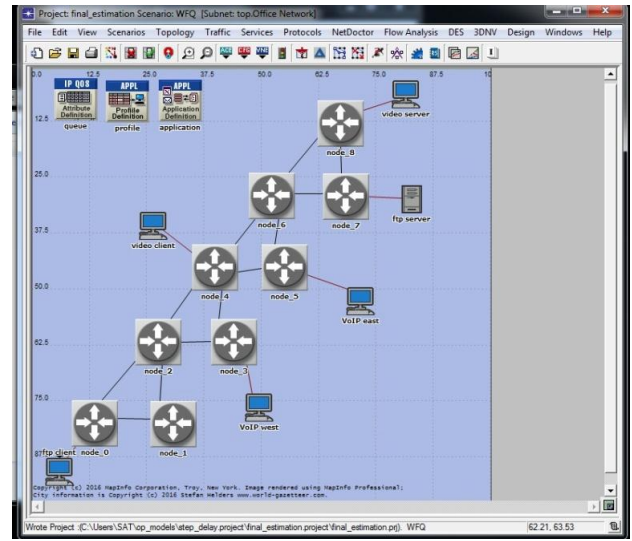

Optimized network engineering (OPNET) software depicts wide varieties of simulations in heterogeneous environment using different protocols. We have used OPNET modeler 16.0 for simulation purpose. OPNET tool simulates the network and performs the powerful functions. Firstly this software was developed for military but nowdays it is used as a network simulation tool as well as a research tool for designing and analysis of networks. The hierarchical structure is divided into three domains as network domain, process domain and node domain. It is also known as event based network simulation tool. The hierarchical structure is divided into three domains as network domain, process domain and node domain. It is also known as event based network simulation tool. Key features of OPNET are:
(i) Modeling and simulation.
(ii) Hierarchal Modeling
(iii) Automatic Simulation Generation
(iv) Implementing and developing new data communication networks.

To simulate performance of various scheduling techniques, the architecture of step network is shown in Table 1.

| Application | ToS | Description |
|---|---|---|
| FTP and Video Servers, FTP and Video Clients and VoIP nodes | Best effort(0) | Connected to routers by 10 BaseT (10 Mbps) links. |
| FTP application | Best effort(0) | High Load, constant (10) for "inter-request time", constant (1,000,000) for "file size". |
| Video application | Multimedia streaming video(4) | Low Resolution Video for video, "Streaming Multimedia" for ToS. |
| VoIP application | Intercativevoice(6) | PCM Quality Speech" for voice, "Interactive Voice. |

**Table 1. List of Events for Transition Table**

*C. Queuing Disciplines*

Packets arrive from different route at intermediate device such as switch and router for processing. Real time traffic such as video conferencing and voice traffic need proper bandwidth with minimal latency, jitter or packet loss as more sensitive towards the network QoS. QoS build for the given networks as:
(1) Adequate bandwidth allocation for all types of traffic i.e. voice, video and data transmitted over a network.
(2) Classification of packets on basis of priority as voice packets gets higher priority compared to video and FTP packet has lower priority.
(3) Queuing occurs in routers and switches by creating different buffer or queue for different types of traffic.

*1. FIFO*

First in first out is a simplest method as in which packets wait in queue until the node is ready to process it. In this technique, first packet arrives is to be first transmitted. In this technique, all the packets have equal opportunity regardless of application or importance of packets as all the packets are treated equally.

*2. PQ*

In Priority queuing packets are processed according to priority The packets having highest priority are processed first and lowest priority packets are processed at last. The packets are sorted in buffer according to priority that reflects the importance and urgency for transmission of packets. PQ is made from different buffers and real time application as VoIP and video conferencing traffic given higher priority so that it can observe minimal delay. But the cons of this scheduling is higher priority packets gets continuous chance to be processed and lower priority packets never processed so starvation will be occur.

*3. WFQ*

A best scheduling technique is in which packets are assigned to different classes and allocated to different queues. But queues are weighted on priority, higher priority queue has higher weight and packets are processed in round robin manner by selecting packets from each queue corresponding to their weight. For example, if weights assigned to a queue are 5,3 and 1 then five packets are processed from first queue, three from second and one from third queue. If priority is not assigned on classes then all weights can be equal, so this technique is also called fair queuing with priority. In weighted Fair queuing packets are weighted based on priority of queue. The system processes packets in each queue in round robin manner and the packets are selected from each queue based on weight.

*D. Performance*

QoS is overall measurement of network performance. The characteristics that measure network performance is bandwidth and can be measured in terms of bandwidth in hertz and bandwidth in bits per second. Throughput is also a measure that actually depicts how fast data is moving on a network. Latency or delay is also a factor that defines the time to send entire message at destination from the first bit sent out from the sender. The latency can be measured in terms of propagation time, transmission time, queuing time and processing delay of data. Jitter is another issue of delay related to real time application. It is delay that occurs packets related to time sensitive application as audio, video data.

*E. UML*

Model is designed to simulate what may or did happen in a real time situation. Modeling is basically done to optimize the real world problem. In presented model a UML class diagram in figure 2 represents structure of a system in general and shows classes, attributes, operations and relationships among classes is designed in which Internet service provider class provides distributed services to the users. The Synchronous Transfer Mode [STM] terminates the fiber optic network. UTP cable is connected to the layer_3 switch that is attached to the server and next layer switch i.e. L2_switch controls and connect the step topology across VLAN and on server final estimation of performance is evaluated.

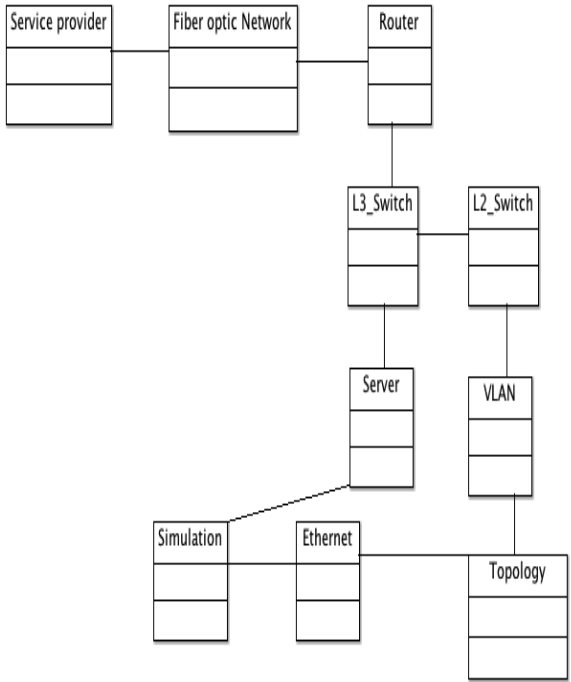

**Figure 2. UML Class Model for Performance Estimation**

IV. EXPERIMENTAL STUDY

### A. Packets Dropped

In case of FIFO, there is only one queue having size of 500 packets and as FIFO processed the packet on first come first serve basis so no priority is assigned to individual type of incoming traffic. When packets arrive more, then they get stored in a queue and if queue becomes full the incoming packets will be dropped.

Another scenario occurs queue becomes full quickly with voice packets then other types of traffic like video and FTP client dropped.

The voice packets and video packets are more dropped in FIFO as packets are treated on First come first served basis and no priority or weight is assigned to packets. Whereas in WFQ packets dropped occur due to congestion state. PQ priority is higher for voice packets than WFQ and FIFO as shown in figure 3.

In case of PQ, Voice packets are assigned highest priority. After voice, video packets get priority then FTP. Voice traffic has supreme priority so they are transmitted as soon as received, and video and FTP traffic has to wait when there is incoming voice packets. Again video traffic has priority over FTP in PQ.

As WFQ handles shared buffer and when buffer gets full congestion arises, now interface enforced queue to be limited. When congestion occurs voice packet has to wait as video packets causing packet loss as queue becomes full which is opposed to PQ where voice traffic has higher priority and need not to be wait.

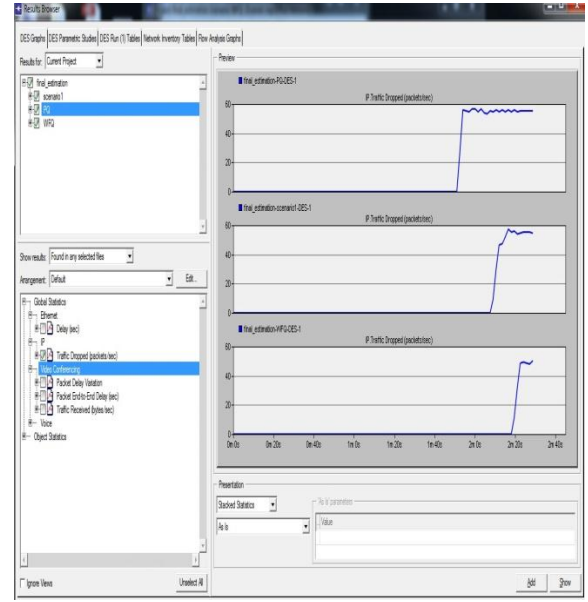

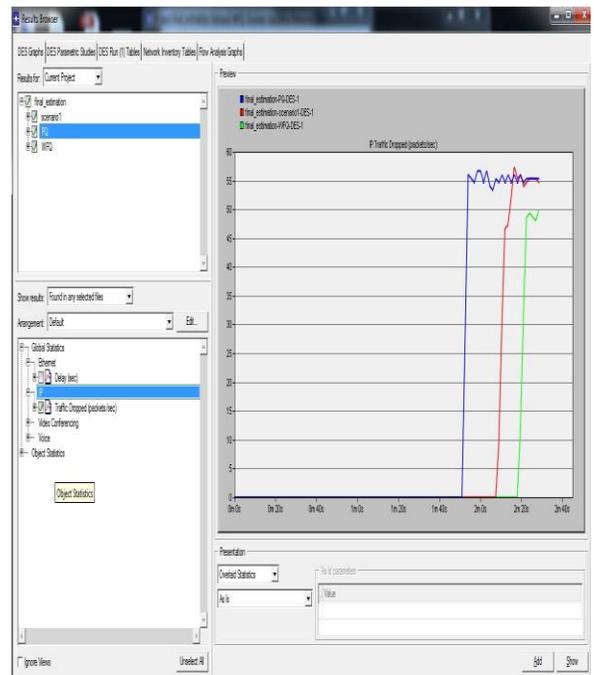

**Figure 3. Packets Dropped**

### B. Delay Variation for Video conferencing Packets:

In case of PQ and WFQ scheduling delay variation is very small because end to end delay is very small for both; also there is no variation in end to end delay because video packets are given somewhat smaller priority in both PQ and WFQ then voice packets as shown in figures 4.. Delay variation is less for PQ and WFQ as compared to FIFO as in FIFO packets are not treated according to priority but allowed on First come First serve basis. PQ and WFQ have the negligible delay variation.

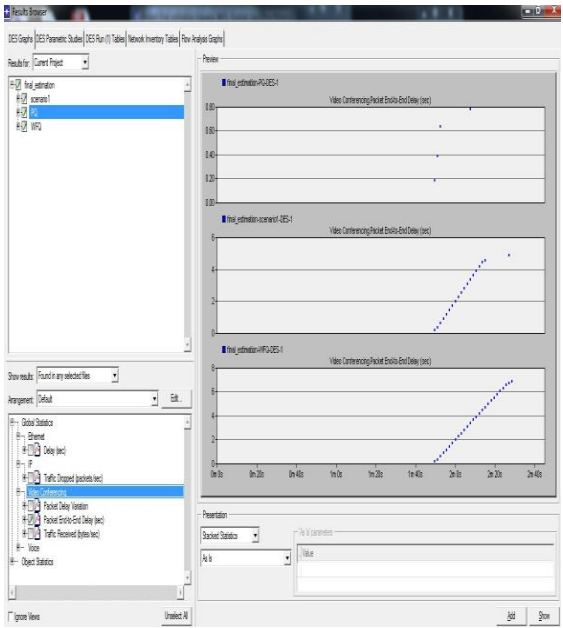

**Figure 4. Delay Variation for Video Conferencing Packets**

*C. Traffic Received for Video Conferencing*

Voice packets occupy in PQ, whole channel because whenever there is voice traffic other traffic is stopped hence traffic received for Video Conferencing is lower in PQ. Now for WFQ, at start traffic received is higher because system is not in congested state and WFQ is using shared buffer and video packets are having lesser weight then voice packets this means video packets are Multipart figures given lesser priority and resulting lesser traffic for video in WFQ as shown in figure 5 and 6.

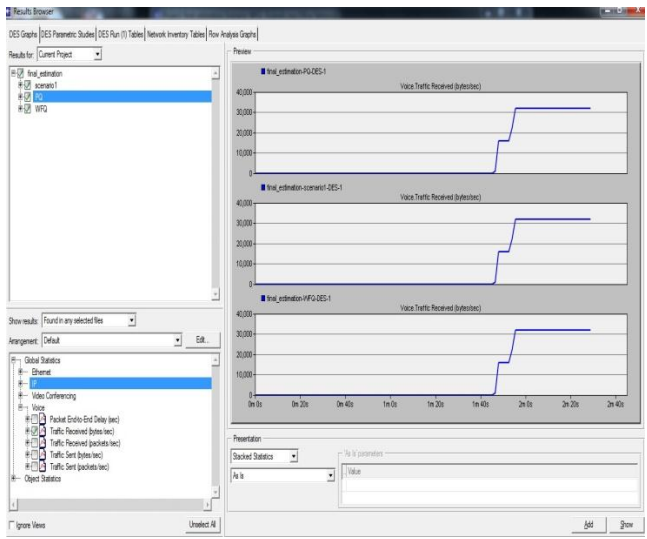

**Figure 5. Traffic Received for VoIP**

*D. Traffic Received for VoIP and Videoconferencing*:

in FIFO Traffic received for voice as compared to the other two scenarios as FIFO implements first come first serve based mechanism, in which no priority is assigned to any type of traffic. Traffic received in case of WFQ decline a little bit due to congestion phenomenon in WFQ. Traffic received for video is less in case of FIFO as compared to WFQ and PQ, also in case of PQ the videoconferencing packets has lesser priority than voice packets and in WFQ at the start the packets are received at higher speed as the network is not in congested state but when congestion occur the traffic received rate will become slower and also video packets has assigned weight 40 while voice packets are given weight 60 this means video packets are given less priority which results in less traffic for video in WFQ.

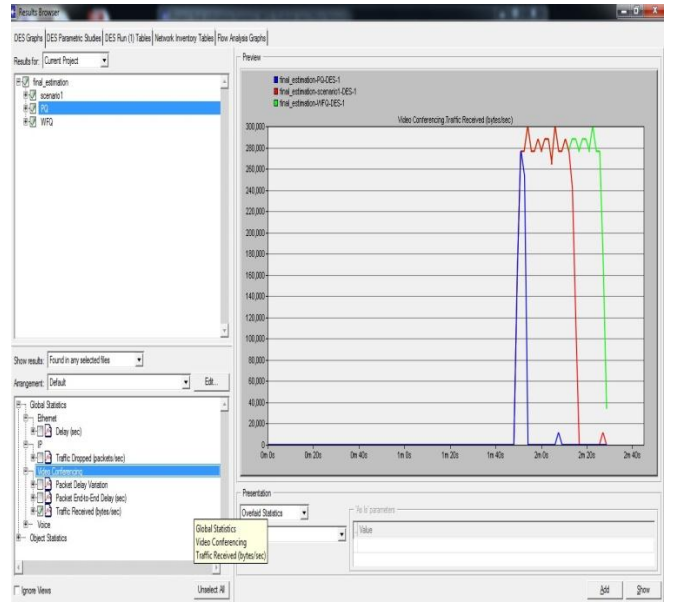

**Figure 6. Traffic Received for VoIP and Videoconferencing**

*E. End-to-End Delay for video conferencing and Voice Packets:*

For PQ and WFQ, the end-to-end delays are almost minor, as maximum priority allocated voice packets as shown in figure 7. For WFQ discipline voice packets have allocated weight 60 and video packets have allocated weight 40 so voice packets has higher weight.

In PQ voice packets are given priority and all other traffic terminated when voice packets incoming and in case of WFQ voice packets have supremacy as higher weight is allocated. In case of FIFO video packets and voice packets has to wait for longer time as packets are treated on FCFS basis.

combining the functionality FIFO and PQ therefore each queue gets turn in round robin fashion. Simulation results illustrate that PQ and WFQ has very small end to end delay and delay variation which is foremost for real time applications i.e VoIP and Videoconferencing. The simulation results also depicted that WFQ and PQ has better performance

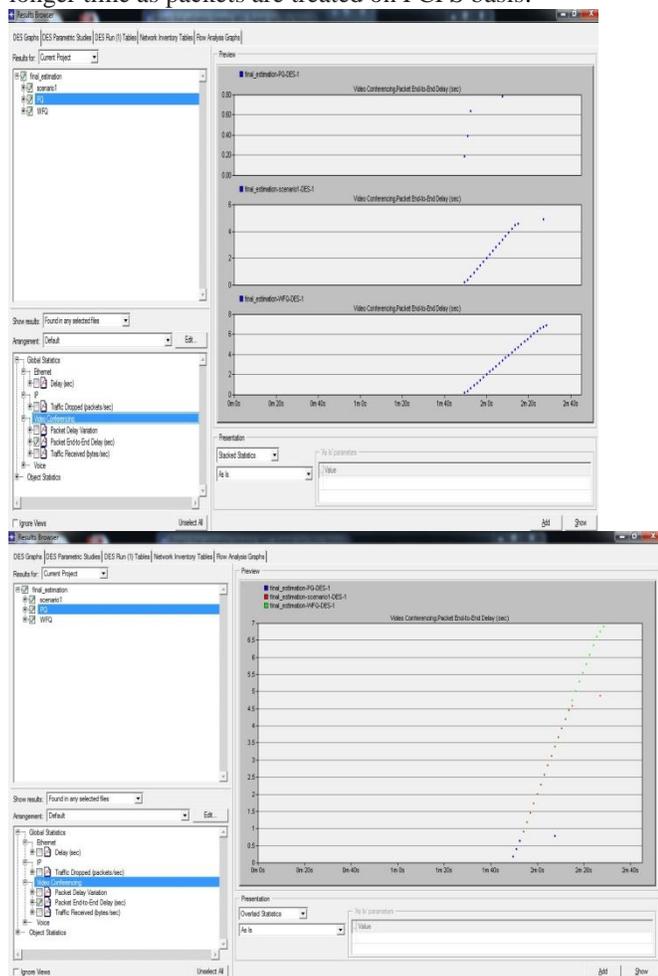

**Figure 7. E-to-E delay for VoIP and Videoconferencing**

V. CONCLUDING REMARKS

The performance of step topology is evaluated and the impact of different queuing scenarios on network is analyzed in terms of packet drops, end to end delay, delay variation, traffic received etc. There is an advantage of using the step topology in comparison of bus topology as link of bus topology fails then it will work in the step network.

FIFO queuing is not best for voice and video packets as it handles the traffic on first come first serve basis so packet drop is more in this scenario where as in PQ voice packets are transferred at higher priority as the priority of voice packets are high compared to video packets. It was analyzed that PQ does not shares bandwidth equally to all other types of traffic so other packets have to wait or rest, WFQ is perform by